\documentclass[12pt]{article}
\def\beq{\begin{equation}}
\def\eeq{\end{equation}}
\def\bea{\begin{eqnarray}}
\def\eea{\end{eqnarray}}
\def\nn{\nonumber}

\begin{document}
\topmargin -0.5cm
\oddsidemargin -0.8cm
\evensidemargin -0.8cm
\pagestyle{empty}
\begin{flushright}
CERN-TH/98-247
\end{flushright}
\begin{center}
\vspace*{5cm}
{\bf SYMMETRY of the RIEMANN OPERATOR}  \\
\vspace*{1cm}
{\bf B. Aneva}  \\
\vspace*{0.3cm}
Theory Division, CERN \\
CH -  1211  Geneva  23 \\
\vspace*{2cm}
{\bf ABSTRACT}    \\   \end{center}
\vspace*{5mm}
\noindent

  Chaos quantization conditions, which relate the
eigenvalues of a Hermitian operator (the Riemann
operator) with the non-trivial zeros of the Riemann
zeta function are considered,
and their geometrical interpretation is discussed.

\vspace*{5cm}
\noindent


\begin{flushleft} CERN-TH/98-247 \\
August 1998
\end{flushleft}
\vfill\eject

\setcounter{page}{1}
\pagestyle{plain}

  For a long time a challenge for mathematical physics has been
and is still the idea, due to Hilbert at the dawn of the
quantum age, to relate the non-trivial zeros of the Riemann zeta
function with a spectrum of a Hermitian operator in a
Hilbert space.

  The Riemann zeta function $\zeta(s)$ is defined for complex
$s = \sigma + i\rho$ and Re$s > 1$ by the equation
\beq
  \zeta(s) = \sum _{1}^{\infty} \frac {1}{n^s}.
\label{1}
\eeq
The Riemann hypothesis \cite{tit} states that the
non-trivial complex zeros of $\zeta(s)$ lie on the
critical line Re$s = 1/2$.

  This is a very good motivation for understanding the
original Riemann conjecture to find a self-adjoint operator
$\partial$ (or a positive one $\partial (1- \partial)$)
defined in a Hilbert space and with a spectrum given by the
non-trivial zeros of the zeta function. In favour of this
idea are the rigorous
results \cite{gelf}, \cite{fad} on the spectral theory of
the Laplace - Beltrami
operator on the Poincar\'{e} complex upper half-plane,
showing that the eigenfunctions are automorphic with
respect to a discrete group of fraction-linear
transformations. The most important application of
representation theory to automorphic forms is the work
of Selberg \cite{sel} who has shown that, if the fundamental
domain of the discrete subgroup $\Gamma \subset SL(2,R)$
is compact, then the spectrum of the elliptic operator
for which the eigenvalue equation
\beq
  -y^2\left(\frac {{\partial}^2}{\partial x^2} +
  \frac {{\partial}^2}{\partial y^2}\right)f =
   \frac {1}{4}(1 + {\rho}^2)f
\label{2}
\eeq
has $\Gamma $-invariant solutions is determined by the
real values $\rho_n$ given by
$\zeta_{\Gamma}(1/2 + i\rho /2) = 0$, where $\zeta_{\Gamma}(s)$
is a meromorphic function of $s$ on the whole plane.
The spectrum is continuous if the fundamental domain has
a finite volume.

  There has been theoretical \cite{b}, \cite{kha}
and experimental \cite{beur}, \cite{od} evidence in support
of the Riemann hypothesis, based on the fluctuations of spacing
between consecutive zeros of zeta.

  In a recent paper, Connes \cite{con} has given a
spectral interpretation of the critical zeros  and
a geometrical interpretation of the explicit formulas
of number theory as a trace formula in a non-commutative space.

  This work was inspired by the of idea of Berry
and Keating, discussed in Cambridge \cite{ber1} and in a
subsequent paper \cite{ber2},
that the real solutions $E_n$ of
\beq
  \zeta(1/2 + iE_n) = 0
\label{3}
\eeq
are energy levels, eigenvalues of a quantum Hermitian
operator (the Riemann operator) associated with the
one-dimensional classical hyperbolic Hamiltonian
\beq
  H_{cl}(x,p) = xp,
\label{4}
\eeq
where $x$ and $p$ are the conjugate coordinate and
momentum. They suggest a quantization condition generating
Riemann zeros, for which however they `see no way to
interpret geometrically'.

  We propose that such a condition can be consistently
obtained as a boundary quantization condition of a
hyperbolic dynamical system within conformal geometry.

  The characteristic properties of the classical dynamics
(the Riemann dynamics) and the quantum analogue
are listed and commented in \cite{ber2}. It has the main
features of a chaotic dynamics, namely instability,
complex periodic orbits, no time-reversal symmetry,
this in turn makes the quantum-classical correspondence
not yet completely understood.

  A prototype for the development of a theory of hyperbolic
systems has been the free motion on a surface of constant
negative curvature (a pseudosphere), which is one of the
first models of chaotic motion. On this surface there
exists a well-defined quantum dynamics where the
Laplace - Beltrami operator acts as the Hamiltonian operator.
The pseudosphere is modelled by the Poincar\'{e} unit disc or
by the complex upper half-plane, which is the conformal
image of the unit disc under a particular fraction-linear
transformation \cite{vor}. The metric is conformal,
proportional to the plane Euclidean metric at each point.
This suggests an analysis of the hyperbolic dynamical system
and its quantum analogue by applying the methods of
conformal representation theory. A motivation for this is
the geometry of points at infinity (or
equivalently all directions along the cone),
namely the property that any isometry of the
hyperbolic space can be extended to a conformal
diffeormorphism of the boundary sphere, which is known
as the `Moebius transform'.

  The classical Hamiltonian $H = xp$ of linear dilation,
i.e. multiplication in $x$ and contraction in $p$, gives
the Hamiltonian equations:
\beq
  \dot{x} = x, \quad\quad\quad\quad \quad
  \quad\quad \dot{p} = -p.
\label{5}
\eeq
Their solutions
\beq
  x(t) = x_0\exp(t), \quad\quad p(t) = p_0\exp(-t)
\label{6}
\eeq
describe, for any $E\neq0$, the classical trajectory
of a hyperbola $E=x_0p_0$. The system is unstable because
it has a hyperbolic fixed point at $x=0, p=0$.

  The system is quantized by considering the dilation
operator in the $x$ space
\beq
  H = \frac {1}{2} (xp + px) =
  -i\hbar\left( x\partial_x + \frac{1}{2}\right),
\label{7}
\eeq
which is the simplest formally Hermitian operator
corresponding to the classical Hamiltonian (4).

  The eigenvalue equation
\beq
  H\psi_E(x) = E\psi_E(x)
\label{8}
\eeq
is satisfied by the eigenfunctions
\beq
  \psi_E(x) = \frac {C}{x^{1/2 - iE/\hbar}},
\label{9}
\eeq
where the complex constant $C$ is arbitrary, since the
solutions are not square-integrable.

  The momentum eigenfunctions are found by the
Fourier transform
\beq
  \phi_E(p) = \frac {1}{\sqrt {h}}
              \int _{-\infty}^{\infty} \psi_E(x)\exp(-ipx/\hbar)dx.
\label{10}
\eeq
To calculate the integral the function $\psi_E(x)$ has to be
analytically continued across the singularity at $x=0$. Following
\cite{ber2}, one might simply choose even eigenfunctions, which
is a natural choice, since the classical hyperbolic
Hamiltonian (4) is parity-symmetric. The
result is (by using the reflection and duplication
formulas for the gamma function):
\bea
  \phi_E(p) &=& \frac {C}{{\vert {p} \vert}^{1/2 + iE/\hbar}
            \sqrt {2\pi}} \int _{-\infty}^{\infty}
    \frac {\exp(-iu)}{{\vert {u}\vert}^{1/2 - iE/\hbar}}du \nn  \\
        &=& \frac {C}{{\vert {p} \vert}^{1/2 + iE/\hbar}}
      (h/{\pi})^{iE/\hbar} \frac {\Gamma(1/4 + iE/2\hbar)}
       {\Gamma(1/4 - iE/2\hbar)}.
\label{11}
\eea

  Thus the quantum Hamiltonian (7) admits two independent
stationary solutions \cite{non} for any real energy $E$
\beq
  \psi_E(\pm {x}) = C\theta(\pm {x})e^{(iE/\hbar - 1/2)log\vert {x}\vert}
\label{12}
\eeq
together with the corresponding momentum eigenfunctions.
In (12) $\theta(x)$ is the Heaviside step function.
In \cite{ber2} a more general relation between coordinate and
momentum eigenfunction is considered; this is interpreted as
two phase-space currents, the $x$ current flowing out from the
origin and the $p$ current flowing into the origin.

  The aim of our study is to obtain a discrete
energy spectrum for the Hamiltonian (7) by imposing some
quantization conditions. Quantization rules may arise
in a natural way by the symmetries of the Hamiltonian (4).
An obvious symmetry is the invariance under scale
transformations:
\beq
  x' = \lambda x ,\quad\quad\quad p' =  \frac {1}{\lambda} p
\label{13}
\eeq
with $\lambda$ positive. It follows from the classical
equations of motion that the parameter of dilation
$\lambda$ corresponds to evolution after time
$T = \log \lambda$. We shall consider the scale invariance
of the hyperbolic dynamical system as
a part of a larger group of transformations
and disscuss the Hamiltonian (4)
from the point of view  of conformal representations
theory. Conformal symmetry in one-dimensional $x$ space
contains translations $x' = x + a$, dilations
$x' = \lambda x$,
and special conformal transformations.
We note that the special conformal transformation
\beq
  x' = \frac{x + cx^2}{1 + 2cx + c^2x^2}
\label{14}
\eeq
can be
written down as a superposition of two conformal inversions
$R$ (i.e. Weyl reflections),
\beq
  Rx = -\frac{1}{x}
\label{15}
\eeq
and a translation $T_{-c}x = x - c$,
\beq
  x' = RT_{-c}R.
\label{16}
\eeq
These transformations are generated by the differential operators
\bea
  P &=& -i\partial_x ,\quad\quad\quad D = -id -ix\partial_x, \nn \\
  K &=& -2idx -ix^2\partial_x,
\label{17}
\eea
which form a realization of the Lie algebra of the conformal
group in one dimension,
isomorphic to the Lie algebra of $SO(2,1)\sim SL(2,R)$; $d$
is a complex number, a conformal (or scale) dimension that
labels the irreducible representations (IR).
We shall briefly review the conformal group in one dimension
$SO(2,1)\sim SL(2,R)$ and
its principal series representation \cite{ser}, which will be
used in the following. The group
$SO(2,1)$ is the group of isometries of the pseudosphere
$$
  -y^2_0 + y^2_1 + y^2_2 = -R^2,
$$
the image of which is the unit disc, by a stereographic projection
onto the plane.  The conformal group in one dimension,
known as the real Moebius group $SO(2,1) = SL(2,R)/Z_2$, acts as
a group of automorphisms of the compactification $S^1$ of the real
space $\cal R$, which can be realized as a set of `light' rays
belonging to the cone $y_0^2 = y_1^2 + y_2^2$. This set of rays
is in a one-to-one correspondence with the unit circle in two
dimensions. The $SO(2,1)$ generators
$X_{ab} = - X_{ba}$, $a,b = 0,1,2$ are expressed in terms
of the conformal group generators by the relations:
\beq
  P = X_{10} + X_{12}, \quad\quad\quad  K = X_{12} - X_{10},
  \quad\quad\quad D = X_{20}.
\label{18}
\eeq
The image of the $R$-inversion (15) on the cone is a reflection
of the axes $y_1,y_2$, i.e. space reflection. In fact the
stability subgroup of the point $x=0$
(or $y_1=0, y_0=y_2=\pm 1$ in homogeneous coordinates) is
isomorphic to the Weyl subgroup. The latter is a subgroup of
$SL(2,R)$ of order $2$ generated ($mod \pm 1$) by a non-trivial
element $w$, with $w^2=-1$.

  The unitary representations of $G = SL(2,R)$ are induced
by an unitary representation of its maximal compact subgroup
$K = SO(2)$ of rotations (generator $X_{12}$). The
representation theory is constructed in the space of functions
with domain in the  homogeneous space $X=K\setminus G$
and
values in a separable Hilbert space. The factor space
$X$ is isomorphic to the upper half-plane. It can be realized
as the space of homogeneous functions of two variables of
given parity and degree of homogeneity, determined by the
conformal dimension $-d$. There is a one-to-one
correspondence between the points of $X$ and the points on the
real line ${\cal R}^1$. An important class of unitary IR is the
so-called  principal series (and its dual) characterized
by the value $-d = -1/2 +i\rho$
(and $-d = -1/2 -i\rho$ respectively), with $\rho$ real.
Let $C^{\infty}$ be the space of infinitely
differentiable (complex-valued) functions $f(x)$ of given
parity, which satisfy certain asymptotic conditions for
$\vert x \vert \rightarrow \infty$ following from the
smoothness of $f$ in the neighbourhood of the $R$-inversion
element.
The unitary (even) principal series IR acts in the
Hilbert space completion $\cal {H}^L$ of $C^{\infty}$
with respect to the scalar product
\beq
  (f_1,f_2) =  \int _{-\infty}^{\infty}\bar{f} _1(x)f_2(x)dx,
\label{19}
\eeq
which is positive-definite and $G$-invariant. One can use
any
$SO(2)$ invariant measure to construct a unitarily
equivalent representation on the upper half-plane.

  There exists a pair of intertwining maps relating the
representation $T^{\chi}, \chi = i\rho$ and the dual
one $T^{\tilde{\chi}}, \tilde{\chi} = -i\rho$. It is defined
by the convolution integral
\beq
  (G_{\chi}f)(x_1) =
      \int _{-\infty}^{\infty}G_{\chi}(x_1-x_2)f(x_2)dx_2,
\label{20}
\eeq
and has the intertwining property
$$
  G_{\chi}T^{\tilde{\chi}}=T^{\chi}G_{\chi}.
$$
The kernel of the integral operator (20) is given by
\beq
  G_{\chi}(x) =
   \frac {N_{\chi}}{\sqrt {2\pi}} {\vert x \vert}^{-1+2i\rho}.
\label{21}
\eeq
The Fourier transform of $G_{\chi}$,
\beq
  G_{\chi}(p)= \int _{-\infty}^{\infty}G_{\chi}(x)e^{-ipx}dx,
\label{22}
\eeq
restricts the choice of the normalization constant $N_{\chi}$
in such a way that, for equivalent representations, the product
$G_{\chi}G_{\tilde{\chi}}$ is proportional to the unit operator
\beq
  G_{\chi}(p)G_{\tilde{\chi}}(p)=1.
\label{23}
\eeq

  The Casimir operator of $SO(2,1)$
$$
  C_2 = -D^2 + \frac {1}{2}(PK + KP) = d(1-d)
$$
corresponds to the Laplace - Beltrami operator $L$
\cite{ser}, which is the second-order differential
operator associated with the $SL(2,R)$-invariant metric
on the upper-half plane $z=x+iy$ and
has the form $L=-y^2 \triangle = -y^2(\partial ^2_x +
\partial ^2_y)$. The solutions of the equation $Lf = d(1-d)f$
are invariant automorphic forms with respect to the discrete
subgroups $\Gamma $ of $SL(2,R)$.  Automorphic forms are
determined by the relations
$f(\gamma z) = f(z)$, for $\gamma \in \Gamma$,
which serve as
boundary conditions for the operator $L$. They define the
Laplace - Beltrami operator as a self-adjoint operator
acting in a
Hilbert space of square integrable functions with respect to
the invariant measure in which a unitary representation of the
group $G$ is realized. The procedure of extending $L$, which is a
positive, semi-bounded and symmetric operator with a dense domain,
to a self-adjoint operator is known \cite{ser}.
For unitary representations containing a stability vector with
respect to the maximal compact subgroup, the spectral decomposition of
$L$ and the associated eigenfunctions - proper and
generalized - are intimately connected with the discrete
subgroup. According to the rigorous results on the $L$-theory
if the discrete subgroup is such that the space
$G/\Gamma $ has a finite volume, the continuous spectrum
entirely lies on the real half-axes $1/4 \leq d(1-d) \leq \infty$,
while the point spectrum lies on the line
$0 \leq d(1-d) \leq \infty$.
The point spectrum is the set of singular points of the kernel
of the resolvant of $L$ for which the  invariant bilinear
functional in the representation space is degenerate.
The automorphic eigenfunctions satisfying $Lf=d(1-d)f$
are generalized functions
and form a complete set of functionals in the space of linear
functionals on the unitary representation Hilbert space.

  The coordinate eigenfunctions $\psi_E(x)$ belong
to the space (denoted hereafter $\Omega ^*$) of linear
functionals on the
even unitary principal series  IR with $\rho = E/\hbar$.
The position functions $\psi_E(\frac{1}{x})$ also belong
to the Hilbert space extension $\Omega ^* \supset \cal {H}^{L}$.

  Returning to the eigenvalue equation (8),
we may interpret it as
a condition for a scale invariance of a conformal wave
function of conformal dimension $d = 1/2 - iE/\hbar$.
If we postulate invariance of the wave function $\psi(x)$
under the full conformal group, this will give a $constant$
eigenfunction. It can be easily verified that in addition
to scale invariance the wave function $\psi(x)$ is invariant under
conformal $R$-inversion:
\beq
  \psi(x) = \frac {1}{x^{2d}}\psi\left( -\frac{1}{x}\right).
\label{24}
\eeq

  Formula (24) defines the coordinate function as
an automorphic form of weight $d$ in the space of the
IR principal series representation, invariant with respect
to the discrete Weyl subgroup of $SL(2,R)$, where its
non-trivial element is chosen to be the conformal $R$-inversion.
(Automorphic is used here in the sense of invariant functions,
associated with a stationary discrete subgroup of automorphisms
of a manifold, which is the homogeneous space of a Lie group.)

  The scale- and $R$-invariance properties of $\psi(x)$ suggest
that the quantum wave function of a chaotic dynamical system (4)
is a scale-invariant function, i.e. a homogeneous function
of degree $-d$ and an $R$-inversion invariant automorphic
function of weight $d$, where $-d$ is the scale dimension
that labels the unitary irreducible  principal series
representation $T^{\chi}$
and is hence equal to $-1/2 +iE/\hbar$, with $E$ real. The
complex-conjugated wave function $\bar{\psi}(x)$ corresponds
to the dual representation $T^{\tilde{\chi}}$
with $-d=-1/2-iE/\hbar$. The above
suggestion means precisely that the wave function of such
a chaotic system is related to the kernel (21) of the
intertwining operator for the IR principal series and its
dual. As such it belongs to the space $\Omega ^*$ of linear
functionals on the unitary principal
series IR Hilbert space. This assumption assures that the
eigenvalues of the formally Hermitian Hamiltonian (7) are real.

  We shall now show that the real values $E$ are eigenvalues of
a self-adjoint operator. For this purpose we consider the
shifted Hamiltonian $1/2 -iH/\hbar = -x\partial_x$,
which satisfies
\beq
   x\partial _x(1- x\partial _x) = -x^2\partial ^2_x.
\label{25}
\eeq
The operator on the RHS of (25) is the Laplace - Beltrami
operator $L$ on the real $x$ line. It is readily seen
that the functions
\beq
  \psi _{\pm E}\left( \frac {1}{\vert x \vert}\right)
  = Cx^{1/2 \pm iE/\hbar}, \quad\quad\quad x > 0,
\label{26}
\eeq
are eigenfunctions of this operator.  We emphasize that the
position eigenfunctions $\psi(x)$ are even functions of $x$.
As seen from the explicit expression (12) they are functions
of the argument  $\vert x \vert$ only and can be considered
as functions defined on the positive real line $(0,\infty)$.
This assumption identifies $\pm x$ and allows the use of
a parametrisation such that the representation space can also
be realized as the Hilbert space of functions $f(x)$,
defined on the half line $x>0$ and square-integrable with
respect to the measure $x^{-2}dx$.
It is known \cite{ser} that the functions
\beq
  \theta (x,d) = x^{1/2 +i\rho}+ c(1/2+i\rho)x^{1/2-i\rho},
\label{27}
\eeq
where $c(d)c(1-d)=1$, with the boundary condition at $a$ with
an additional parameter $\kappa$,
\beq
  \theta(a) = a\kappa \theta'(a),
\label{28}
\eeq
form a complete set
of eigenfunctionals  of the self-adjoint operator $L$ in the
space of linear functionals on the unitary principal series
IR  Hilbert space of functions $f(x)$ defined on the positive line
$a\leq x \leq \infty, a \geq 1$ and square-integrable with respect
to the measure $x^{-2}dx$.

  Let $\psi_{\pm E}(1/x)$ be the eigenfunctions
of the Laplace - Beltrami operator $-x^2\partial_x$
corresponding to the eigenvalues $1/4 + E^2/{\hbar}^2$ of
the continuous spectrum of $L$.
It follows from (25) that
\beq
  x\partial_x(1 - x\partial_x)x^{1/2 \pm iE/\hbar} =
  \left( \frac {1}{4} + \frac {E^2}{{\hbar}^2}\right)
  x^{1/2 \pm iE/\hbar}.
\label{29}
\eeq
It is clear that $x^{1/2 \pm iE/\hbar}$ are eigenvectors of
the operators $x\partial_x$ and $1-x\partial_x$, corresponding
to the eigenvalues $1/2 \pm iE/\hbar$ and $1/2 \mp iE/\hbar$
respectively. Thus if $1/4 + E^2/{\hbar}^2$ belongs to the
spectrum of $L$ then $1/2 \pm iE/\hbar$ belongs to the spectrum of
the dilation operator $x\partial_x$ and vice versa, so that
roughly speaking the dilation operator is the square root of the
Laplace - Beltrami operator in the Hilbert space of functions
with domain  $(a, \infty)$ and square integrable with respect
to the measure $x^{-2}dx$.

  We turn now to the eigenvectors $\psi(x)$. As can be
verified immediately they are eigenvectors of $L$ but
corresponding to complex eigenvalues
$1/4 - (1\pm iE/\hbar)^2$. However, there is an operator
which, when acting on the position eigenfunctions $\psi(x)$,
yields the real eigenvalues  $1/4 + E^2/\hbar^2$.
It is the $R$-transformed Laplace - Beltrami operator
denoted hereafter  $L_R$. It has the form
\bea
   L_R &=& -x\partial_x(1 + x\partial_x) \nn  \\
       &=& -x^2\partial ^2_x -2x\partial_x.
\label{30}
\eea
Expressing $L_R$ in terms of the Hamiltonian (7):
\beq
   L_R = (1/2 - iH/\hbar)(1/2 + iH/\hbar)
\label{31}
\eeq
we obtain that the position functions $\psi(x)$ are
eigenfunctions of $L_R$ corresponding to the
eigenvalues $1/4 + E^2/{\hbar}^2$, so that the
operators $1/2 \pm iH/\hbar$ are formally the square roots
of a positive operator. The elliptic Laplace - Beltrami
operator and the $R$-transformed one act on functions
whose arguments are related by the $R$ transformation.
The Laplace - Beltrami operator is densely defined and
self-adjoint on the line $[1, \infty)$. The continuous
$R$-transformation $x'= 1/x$ on the line $x>0$,
maps its eigenfunctions to the eigenfunctions
of the $R$-transformed operator $L_R$, which is hence
densely defined and self-adjoint on the interval $(0, 1]$.
The domains of the operators consist of elements
satisfying the relation for
automorphic $R$-invariant functions and have as common
elements the functions $f$ defined for $x=1$, which is
the fixed point of the transformation $x'= 1/x$.
The two operators have to be equal on the common elements.
Relations (24), which determine
the eigenfunctions $\psi(x)$ as invariant automorphic
functions with respect to the conformal $R$-inversion,
serve as boundary conditions  at the point $x=1$ (or at any
point $a \geq 1$, if $x'= a/x$), for coincidence of the inversed
operator $L_R = (1-iH/\hbar)(1+iH/\hbar)$ with the self-adjoint
Laplace - Beltrami operator
and will be used in phase space as
quantization conditions that generate a
discrete spectrum. Thus due to the $\pm x$ identification
the $R$-invariance automorphic relation defines the
dilation Hamiltonian as a square-root of a self-adjoint
operator.
For the eigenfunctions of the Laplace - Beltrami operator,
eq. (24) is equivalent to the
Neumann boundary conditions for the functions and for their
first derivatives:
\beq
  \psi \left( \frac {1}{\vert x \vert}\right) =
  \psi \left( \frac {1}{\vert x \vert}\right),
  \quad\quad\quad \psi '\left( \frac {1}{\vert x \vert}\right) =
   - \psi '\left( \frac {1}{\vert x \vert}\right),
\label{32}
\eeq
at the point $\vert x \vert =1$. Cleary thanks to the
additional paramater in (28) these conditions are trivially
fulfilled. If not only the first of these conditions is
automatically satisfied, while the second one has only
the trivial solution, unless it is assumed that
$\vert x \vert = 1$ is a point where orientation is
reversed. This seems to be a natural assumption, since
conformal $R$-inversion is an orientation-reversing
transformation, with fixed points $\vert x \vert =1$, so
that the points $x$ may belong to manifolds of opposite
orientation. Such are the
curves of constant energy in phase space. The volume
element in phase space is given by the area $xp=h$ of the
Plank cell with sides $l_x$ and $l_p$.
Contours of constant energy in phase space are
the upper and lower branches of the hyperbola $E=px$ and
of the conjugated one $E=-px$, corresponding to the
eigenvalues of the of the square root operator $\pm H$
of the $R$-transformed Laplace - Beltrami operator. Being
aware of the time non-invariance of the classical orbits
\cite{ber2}, one should interpret the upper and lower branches
of the conjugated hyperbola respectively as the
combined space-time transform and the time transform of a
given one. The $R$-inversions in the $x$-space are easily
enlarged to transformations in phase space, which are
area preserving and leave invariant the operators $\pm H$.
These transformations have the form:
\bea
  T_1^{\pm}:\quad\quad\quad x'&=&-\frac {h}{x}\quad\quad\quad
  p'=\pm \frac {x^2p}{h}, \nn \\
  T_2^{\pm}:\quad\quad\quad x'&=&-\frac {h}{p}\quad\quad\quad
  p'=\mp \frac{xp^2}{h}.
\label{33}
\eea
The upper $\pm$-indices denote that the determinant of the
transformation is $\pm 1$.  The $-T_1$ transform of $x$
is the conformal $R$-inversion corresponding to the value
$a=2\pi$ in (28) (in units $\hbar=1$). For $xp=h$ the
transformations contain the canonical one $x'=-p , p'=x$
and reflections $x'=\pm x, p'=\pm p$. Together with the
transformations $-T_1^{\pm}$ and $-T_2^{\pm}$ (note that
the identity is $-T^+_2$), they generate a finite group
which is the dihedral group ${\cal D}_4$ of order $8$
\cite{gro}, consisting of all orthogonal transformations in
two dimensions that preserve the regular $4$-sided polygon
centered at the origin. It is actually generated by
reflections (denoted hereafter $S_1$ and $S_2$ respectively)
about the $x$ axis and about the (diagonal) line $l$
inclined at an angle $\pi /4$ to the positive $x$-axis.
The rotations through multiples of $2\pi /4$
form a cyclic subgroup of order $4$. The invariant $4$-gon
is spanned by the vectors
$(\pm \sqrt h,0), (0,\pm \sqrt h), (\pm \sqrt h,\pm \sqrt h)$.
A fundamental region in the square is the $2$-simplex with
vertices $(0,0), (\sqrt h,0), (\sqrt h,\sqrt h)$.
The $2$-simplices are
joint two by two to form four $2$-complexes (the Plank cells
in the distinct quadrants). The transformations fall into
two classes depending on whether they preserve (reverse)
sign of volume element, i.e. preserve (reverse)
orientation. We can use the isometries that map one side
of the fundamental region onto a side of an adjacent region
to glue distinct regions together along their boundaries,
imposing boundary and quantization conditions and forming
a closed path corresponding to a classical closed orbit.
By identifying all boundary points mapped to one another
by an element of the group (i.e. points in each orbit of
the group) we can impose as a boundary condition the
requirement that the coordinate and the momentum eigenfunctions
are automorphic invariant forms of the discrete subgroup of
reflections:
\beq
  \psi(x)= \psi(sx),\quad\quad\quad\phi(p)=\phi(sp),
\label{34}
\eeq
for all $s$ in the group. This is possible since for $xp=\pm h$,
the $x$ and $p$ transformations each form a one dimensional
factor map $x'=T(x)$, $p'=T(p)$ of the two dimensional map (33)
of the $4$-gon.
The reflections in the $x$-factor map are the conformal
$R$-transformation and the $x$-inversion and generate the Weyl
subgroup of the dihedral group. The coordinate function $\psi(x)$
is thus defined to be an invariant automorphic function with respect
to the discrete $R$-transformation (which it is) and
eq. (34) is equivalent to
\beq
  \psi(x)=\frac {1}{x^{2d}}\psi(Rx), \quad\quad\quad
  \psi(x)=\psi\left( \pm \frac {h}{p}\right),\quad\quad\quad
  \phi(p)=\phi\left (\pm \frac {h}{x}\right).
\label{35}
\eeq
Eqs. (35) impose relations between values of $\psi$ and $\phi$ at
boundary points mapped into each other by an element $s$,
generating the finite Weyl subgroup of the dihedral group.
Since under the discrete $R$-transformation points transform to
$E>h$ with decreasing $x$ and to $E<-h$ with increasing $x$
and taking the limit $h \rightarrow 0$ yields the semiclassical
approximatiom, it will be thus imposed  as a boundary condition
for a smooth transition from $E<-h$ to $E>h$.
  An orientation is then assigned to each
$2$-cell and a direction to each edge, the natural direction,
due to the defining homeomorphism to the unit interval
$[0,1]$, being from initial point $0$ to final point $1$.
To each distinct element $g_i, i=1,2,..8$ (i.e. root vector)
of the group a vertex $i$ is assigned and a directed edge
$(i,j)$ connects two vertices iff $g_i=S_kg_j$ for a
generator $S_k$ of the group. A path
extends from a given $g$ at an edge of the fundamental simplex
to $S_{i_1}...S_{i_k}g$ and is
closed if it can be presented in the form $S_{i_1}...S_{i_k} = 1$.
This is a consequence of the relation $(S_iS_j)^{m(i,j)}=1$,
with $m(i,i)=1$,
for the Coxeter element \cite{hum} of the dihedral group,
which is just the product of two genetating reflections,
hence is a rotation through $2\pi /4$ of order $4$.
Denoting the rotation through
$\pi /2$ by  $r=S_1S_2$, we obtain that a closed path
$S_1S_2S_1S_2$ is always of the form $rr$,
which is the word representation of a projective plane. Thus
the relation for the Coxeter element forces identification of
antipodal boundary points and introduces topology of
a projective plane \cite{fom} in phase space.
At both $l_x$ and $l_p$ boundary sides of the $4$-gon
$\pm x$ and $\pm p$ are identified,
the latter according to the equivalence
relation $(-\sqrt h, p)\simeq (\sqrt h, -p)$, which
endows the phase space with Moebius topology.
It is readily seen that the Weyl reflection $S_2$  and
$S_2^2$ form a cyclic subgroup of order $2$, isomorphic
to the group $Z_2$. With the identification of boundary points
a closed path begins from a given (initial) point $g_i=(x,p)$
and ends at a (final) antipodal one $g_f=(-x,-p)$. The closed
paths $\gamma_s$ fall into two homotopy classes depending on
whether the final point is reached via a product of even or
odd number of
reflections $S_2$. In the first case, the end point is connected
to the initial one by a transformation with $det=1$ and
orientation is preserved along the path, while in the second
case, the determinant of the transformation
is $-1$ and orientation is reversed along the path. Hence the
two homotopy classes form a representation of the fundamental
group of the projective plane $\pi _1({\cal{RP}}^2)$, which is
the non-trivial homology group $Z_2$.
The boundary conditions for the automorphic functions have
to be compatible with the topology of the projective plane, and
in particular with the Moebius topology equivalence relation.
Due to the twist the original $2$-complexes (the Plank cells in
the original quadrants) are mixed and eq. (35) becomes
\beq
  \psi(x)\vert _{xp=h} = \frac {1}{x^{2d}}\psi\left( -\frac{1}{x}\right)
  \vert _{(-x)(-p)=h} = \frac {1}{x^{2d}}\psi\left( \frac{1}{x}\right)
  \vert _{x(-p)=h}
\label{36}
\eeq
The boundary conditions in the form of eq. (36) are
compatible
with the identification of $\pm x$, and of $\pm p$ by a twist,
and serve at the same time as quantization conditions
generating discrete spectrum.
Indeed we have:
\bea
  \psi(x)&=& \frac {1}{x^{1 - 2iE/\hbar}}
  \frac {C}{x^{-1/2 + iE/\hbar}}  \nn  \\
         &=& \frac {1}{x} \frac {C}{x^{-1/2 - iE/\hbar}}.
\label{37}
\eea
Using further that $x = h/\pm p$ and the expression (11) for the
momentum eigenfunctions, we obtain
\beq
  \psi(x) = \pm \frac {1}{x^{1/2}} p^{1/2} \phi(p) \pi^{iE/\hbar}
                    \frac {\Gamma(1/4 - iE/2\hbar)}
                  {\Gamma(1/4 + iE/2\hbar)}.
\label{38}
\eeq

With the help of the functional equation for the Riemann
zeta function
\beq
  \zeta(s) = \pi^{s-1/2} \frac {\Gamma((1-s)/2)}{\Gamma(s/2)}
                               \zeta(1-s)
\label{39}
\eeq
for $s = 1/2 + iE/\hbar$, we get from (38) the result
\beq
  x^{1/2}\zeta(1/2 - iE)\psi(x) = \pm p^{1/2}\zeta(1/2 + iE/\hbar)
    \phi(p), \quad\quad\quad  x(\pm p) = h.
\label{40}
\eeq
The latter formulas, with $\pm x$ and $\pm p$ identified,
are consistent
only if either $\psi(x)=0$ (and hence $\phi(p)=0$), or
\beq
  \zeta(1/2 + iE/\hbar) = \zeta(1/2 - iE/\hbar) = 0,
\label{41}
\eeq
which provides a regularization procedure
for semiclassical approximation and generates a discrete
real spectrum of the dilation Hamiltonian operator (7).
Geometrically the conditions reflect the fact that the
fundamental group of the projective plane is non-trivial,
and can be mapped homomorphically to the
automorphisms group of order two of the covering space.
The defining homomorphism
$$
  \sigma : \pi_1 ({\cal {RP}}^2,x) \rightarrow \pm 1 \simeq Z_2
$$
assigns to each path $\gamma_s$ a number
$\sigma (\gamma_s) =\pm 1$, according to whether orientation is
preserved or reversed by transport around it in such a way,
that $\sigma (\gamma_s) \neq 1$, for $\gamma_s \neq 1$.
This is consistent with the group structure of the
transformations (33),  which form the double point dihedral
group with the factorization property
${\cal D}_4 = {\cal D}_2\times Z_2$ (it can be presented
equivalently as a semidirect product of the two cyclic
groups $Z_4\cdot Z_2$)\cite{bov}. Hence the
irreducible representations have a $Z_2$-grading too.
They are twice as many, and given a representation $\psi(x)$ of
${\cal D}_2$ each element of the pair
$\sigma ({\gamma}_{s})\psi(x)$
could be interpreted as associated with an orientation
preserving or reversing homotopy
class. For $x$ belonging to a closed trajectory
the quantization conditions
(40) are an algebraic expression
of the homomorphism $\sigma$, this unique homomorphism
being the geometric representation of the finite
Weyl (orthogonal) reflection  group.

  Employing the notation as in \cite{ber2}
\beq
  \langle x\vert \psi\rangle \equiv \psi(x), \quad\quad\quad
  \langle p\vert \psi\rangle \equiv \phi(p), \quad\quad\quad
  \hat{H}=H/\hbar,
\label{42}
\eeq
we rewrite expression (40) in the form:
\bea
  \langle x\vert {\hat{x}}^{1/2}\zeta(1/2 - i\hat{H})\vert \psi_E\rangle
  =\pm \langle p\vert {\hat{p}}^{1/2}\zeta(1/2 +i\hat{H})\vert \psi_E\rangle,
  \nn \\
    xp = h, \quad\quad\quad (-\sqrt h, p) \simeq (\sqrt h, -p).
\label{43}
\eea

  These relations can be geometrically  interpreted as
quantization conditions generating the Riemann zeros
as a discrete spectrum of the
energy operator of the conformally invariant Riemannian
metric due to dihedral symmetry in phase space of
the chaotic dynamical system.
The Riemann zeros are the
eigenvalues of the hyperbolic chaotic Hamiltonian
for which the
latter is the square root of the (inversed)
Laplace - Beltrami operator.

  The relation with the $-$ sign on the RHS of (43) is
exactly the quantization condition suggested by Berry and
Keating in \cite{ber2}.

  Remark: It is known that the surface symbol is not
unique. Hence we can interpret the closed path
either as a cylinder $S_1S_2S_1^{-1}S_2$,
which is obtained from the $4$-gon by
identification of the $l_x$-sides only, or as a Moebius band
$S_1S_2S_1S_2$
with the $l_p$-sides identified by a twist. In the first case
the boundary condition is the trivial one (the $+$ sign
in (43)) and is not a quantization condition. In the case of
a Moebius band the boundary condition is the non-trivial one
(the $-$ sign in (43)) and serves as a quantizatin
condition which generates Riemann zeros.

  To summarize we have implemented  conformal symmetry to
model a chaotic dynamical system. The  quantum Hamiltonian
has a classical limit and a discrete real spectrum whose
points are the Riemann zeros on the critical line
Re$s = 1/2$. The coordinate
eigenfunctions are automorphic functions, invariant
with respect to the discrete subgroup of Weyl reflections,
for points on hyperbolic trajectories of constant energy in
phase space. The smooth semiclassical behaviour around
the saddle point in phase space is only possible for
eigenvalues of the Hamiltonian, which are the critical
zeroes of the Riemann zeta function.

  The author is grateful for the hospitality of the
Isaac Newton Institute in Cambridge and for the possibility
to meet M.Berry and J.Keating there.
The author is also grateful for the support and
hospitality of the Theory Division at CERN where this work
was completed.
Partial support of the Bulgarian
National Science Foundation under contract $\phi -827$ is
acknowledged.

\end{document}